\documentstyle[12pt,epsf]{article}
\newcommand{\be}{\begin{equation}}
\newcommand{\ee}{\end{equation}}
\newcommand{\bear}{\begin{eqnarray}}
\newcommand{\ear}{\end{eqnarray}}

\newsavebox{\LSIM}
\sbox{\LSIM}{\raisebox{-1ex}{$\ \stackrel{\textstyle<}{\sim}\ $}}
\newcommand{\lsim}{\usebox{\LSIM}}
\newsavebox{\GSIM}
\sbox{\GSIM}{\raisebox{-1ex}{$\ \stackrel{\textstyle>}{\sim}\ $}}
\newcommand{\gsim}{\usebox{\GSIM}}

\begin{document}
\begin{titlepage}
\begin{flushright}
HD-THEP-98-48\\
hep-ph/9809506
\end{flushright}
$\mbox{ }$
\vspace{1cm}
\begin{center}
\vspace{.5cm}
{\bf\LARGE SUSY Variants}\\
\vspace{.3cm}
{\bf\LARGE of the Electroweak Phase Transition\footnote{Based
on a talk presented by M. G. Schmidt at the First European
Meeting ``From the Planck scale to the electroweak scale''
(Kasimierz, Poland, May 1998)}}\\
\vspace{2cm}
Stephan J. Huber\\
Michael G. Schmidt\\
\vspace{1cm}
Institut f\"ur Theoretische Physik\\
Universit\"at Heidelberg\\
Philosophenweg 16, D-69120 Heidelberg, FRG\\
\end{center}
\bigskip\noindent
\vspace{3cm}
\begin{abstract}
The MSSM with a light right-handed stop and supersymmetric 
models with a singlet whose vev is comparable to that
of the $SU(2)_W$ Higgs allow for a strongly first-order
electroweak phase transition even for a mass of the lightest Higgs
around 100 GeV. After a short review of the standard model situation
we discuss these supersymmetric models. We also compare perturbative
calculations based on the dimensionally reduced 3-dimensional
action with lattice results and present an analytic
procedure based on an analogue of the stochastic vacuum model 
of QCD to estimate the nonperturbative contributions.
\end{abstract}
\end{titlepage}

\section{Introduction}
The unification of fundamental interactions at high energies is usually 
discussed in the framework of the renormalization group based on
perturbation theory for temperature $T=0$ quantum field theory.
However,  according to our present understanding, high energies much 
above the reach of today's accelerators
were realized in the hot early universe: during its expansion
it cooled down and -- like in material physics (liquid--vapor,
alloys, superconductors...) underwent phase transitions. The latter depend
crucially on the particle content and the interactions of
the underlying theory and on the time scale of the expanding universe.
Thus the study of possible relics of such phase transitions might
reveal interesting news about the basic theory. In general,
these are genuinely nonperturbative phenomena. Therefore one needs
methods to treat them appropriately.

The electroweak standard model (SM) is so successful because
it allows for very accurate perturbative calculations of high
energy processes due to the small weak coupling $g_W$ in the Higgs
ground state and because these agree beautifully with experiments.
Still it is not considered as a fundamental theory above the 1 TeV
scale because it is not stable against impact from GUT-scale physics.
Thus any hint towards a modification of the electroweak standard
model is highly welcome to theorists.

\begin{figure}[b] 
\caption{}
\begin{picture}(200,70)
\put(30,0){\epsfxsize13cm \epsffile{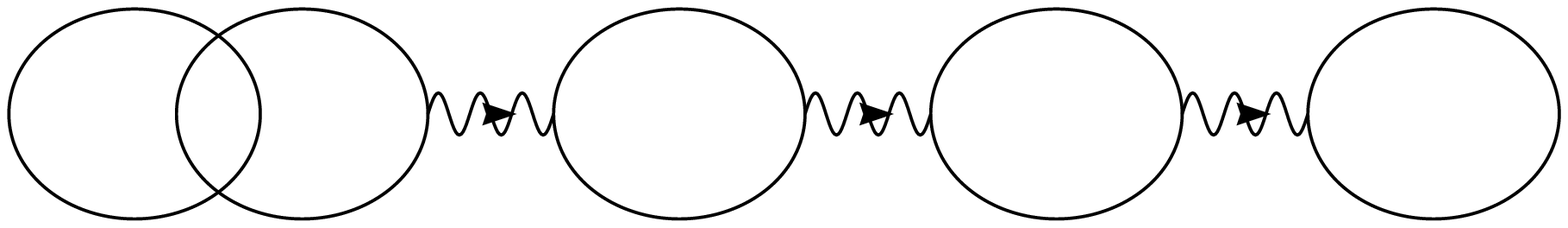}}
\put(40,30){\footnotesize{Inflation}}
\put(35,-5){ \scriptsize{$B=L=0$}}
\put(100,27){GUT}
\put(88,-5){\scriptsize{$B-L\neq 0$} ?}
\put(182,40){\footnotesize{Hot}}
\put(170,25){\footnotesize{Sphaleron}}
\put(170,0){ \scriptsize{$B=L=0$}}
\put(167,-11){ \scriptsize{if $B-L=0$}}
\put(254,40){\scriptsize{electroweak}}
\put(253,25){\scriptsize{baryogenesis}}
\put(255,-5){ \scriptsize{$B+L\neq 0$}}
\put(335,30){ \footnotesize{$\frac{v(T_c)}{T_c}\gsim 1$}}
\put(339,0){ \scriptsize{freeze out}}
\put(343,-10){ \scriptsize{of $B$, $L$}}
%\put(190,-5){Figure 1}
\end{picture} 
\label{fig1} 
\end{figure}

We will first review shortly the well understood phase transition
(PT) in the SM in chapter 2. It has all the necessary ingredients 
postulated by Sakharov to possibly produce  \cite{1,2}
the baryon asymmetry of the universe
$\left((n_B-n_{\bar B})/n_\gamma\sim 10^{-10}\right)$, and this
is the main promise of a standard model PT. However, it turned out
to be not fulfilled. Thus in the discussion of baryogenesis one
might have to return to the old (or new)
out-of-equilibrium decay scenarios
\cite{3} of very heavy particles
contained in GU models. However, grand
unification should be dealt with in close relation to inflation:
inflation is the well-known solution to some problems in standard
cosmology and also provides  a very successful explanation of primordial
density fluctuations in the early universe. Since all particle densities
are diluted exponentially during the inflationary period, the
GUT particles supposed to create some baryon asymmetry have to be produced
in the pre-reheating period after inflation or in a separate
GUT phase transition after inflation. The latter is very problematic
because the reheating temperature after inflation is favourably
not in the range of GUT energies, in particular in SUSY models.
Furthermore, the GU theory has to be $B$--$L$ violating, otherwise
B and L are washed out by the $B$--$L$ conserving (hot) sphaleron
transitions during the thermodynamic equilibrium period before
the electroweak PT (fig.~\ref{fig1}).

Chapter 3 contains some remarks about a (semi)analytic
treatment of nonperturbative effects in a first-order PT based
on recent work of the authors. 
The most important point is an instability of the gauge field $F^2=0$
vacuum of the hot electroweak theory for small Higgs vevs. This 
leads to nonperturbative modifications of the electroweak potential. 
In chapter 4 and 5 we discuss
variants of the SM: in chapter 4 the minimal supersymmetric standard
model (MSSM) with a ``light'' $stop_R$ superpartner of the
right-handed top.
This will allow a strongly first-order phase transition even at Higgs
masses as large as 100 GeV.  
In chapter 5 we consider a nonminimal (next to minimal?)
supersymmetric model (NMSSM) with a singlet superfield $S$ which
also obtains a vev $<S>$ of the order of the Higgs vev.
Different from the standard situation this model leads to
a first-order phase transition already at the tree level. Again Higgs masses
of 100 GeV are compatible with a strong PT in a broad range of parameters.    

\section{Electroweak Phase Transition in the Standard Model}
\setcounter{equation}{0}
The order and the strength of the electroweak PT can be
discussed with the (thermal equilibrium) effective potential
$V_{eff}(\varphi^2,T)$ (free energy) depending on the Higgs background
field $\varphi(<\phi>={\varphi \choose 0}/\sqrt 2)$ and the
temperature. A simple 1-gauge field loop calculation in thermal field
theory results in a positive (Debye) mass contribution
to the Higgs mass proportional to $T$. This reduces the
Higgs instability and predicts a phase transition at  high
temperatures \cite{4}. The Matsubara zero modes (without time dependence)
in the same 1-loop calculation produce a term $\sim -T(\varphi^2)
^{3/2}$ in the potential which leads to two degenerate minima
at some critical temperature $T_c$ labelled by ``symmetric'' $(\varphi
=0)$ and ``Higgs'' $(\varphi_{Min}\not=0)$.
This would naively imply a first-order PT \cite{5}. However, a concise
analysis of the effective potential reveals that this is a very
preliminary result because one has to treat the infrared (IR)
behavior for $\varphi\to0$ properly. In high-temperature
gauge theory the coupling $g_W^2T$ has a dimension, and the
dimensionless ratio $g_W^2/m_{IR}$ is not small if the
scale $m_{IR}\sim \varphi$ becomes small. Furthermore it turns
out that 2-loop contributions are quantitatively very important.
A clean way \cite{7,8} to deal with such a situation
is a stepwise procedure
(``dimensional reduction''): Integrate out\\
1) $n\not=0$ Matsubara
modes with $p_{0_n}=2\pi nT$ (including all fermions with
$n=1/2, 3/2,...$).\\
2) $n=0$ modes of longitudinal gauge fields $A_0$ which have
obtained a Debye mass $m_D\sim g_WT$ in the first step.

Here ``integrate out'' is understood in the sense of a matching
procedure, matching a set of static
4-dimensional amplitudes containing the
above modes in the loops to a 3-dimensional
truncated Lagrangian \cite {9a}
for the Higgs and transversal gauge zero modes:
\be\label{2.1}
L^{3-dim}_{eff}=\frac{1}{4}(F^{a}_{ik})^2+(D_i\phi_3)^+(D_i\phi_3)+m^2_3
(T)\phi^+_3\phi_3+\lambda_3(T)(\phi_3^+\phi_3)^2\mbox{ }.\ee
$L^{3-dim}_{eff}$ contains a 3-dimensional gauge coupling
$g^2_3=g_W^2T(1+...)$, a $T$-dependent Higgs mass $m^2_3(T)$ and
a coupling $\lambda_3(T)=\lambda_T^{4-dim}T$ between 
3-dimensional Higgs fields (canonical dimension 1/2).
Neglection of higher terms in (\ref{2.1}) (e.g. $\sim(\phi_3^+\phi_3)^3$)
introduces a (few percent) error of ${\cal O} (g_W^3$) for $\varphi_3
<2\pi\sqrt T$. It is important to note that steps 1 and 2 above can
be performed in (two-loop) perturbation theory, whereas
the zero-mode Lagrangian (\ref{2.1}) contains all the IR problems for
$\varphi\to 0$. As it stands $L^{3-dim}_{eff}$ contains a potential
which still naively describes a second-order PT at $T_c$ with
$m^2_3(T_c)=0$. However, it is just the tree Lagrangian of
the 3-dimensional theory. Calculating naively again the 1-loop
action in this theory one reproduces the $-T(g^2_W\varphi^2)^
{3/2}$ term mentioned above (now in the form $-(g^2_3\varphi^2_3)
^{3/2})$ leading to a first-order PT. However, we expect important nonperturbative
IR effects if the perturbative 3-dimensional potential competes
with a nonperturbative part at $\varphi_3=0$ which is of order
$(g^2_3)^3$ for dimensional reasons. Rescaling by
powers of $g^2_3$ -- the unique scale in the problem -- we have
the dimensionless couplings
\be\label{2.2}
y=\frac{m^2_3(T)}{(g^2_3)^2}\quad,\quad\quad x=\frac{\lambda_3(T)}{g^2_3}
\left(\sim\frac{\lambda_T^{4-dim}}{g_W^2}\right)\ee
where $y$ is related to $(T-T_c)$ and $x$ is the
parameter  which determines the critical behaviour of (\ref{2.1}).

\begin{figure}[t] 
\begin{picture}(200,165)
\put(60,-10){\epsfxsize10cm \epsffile{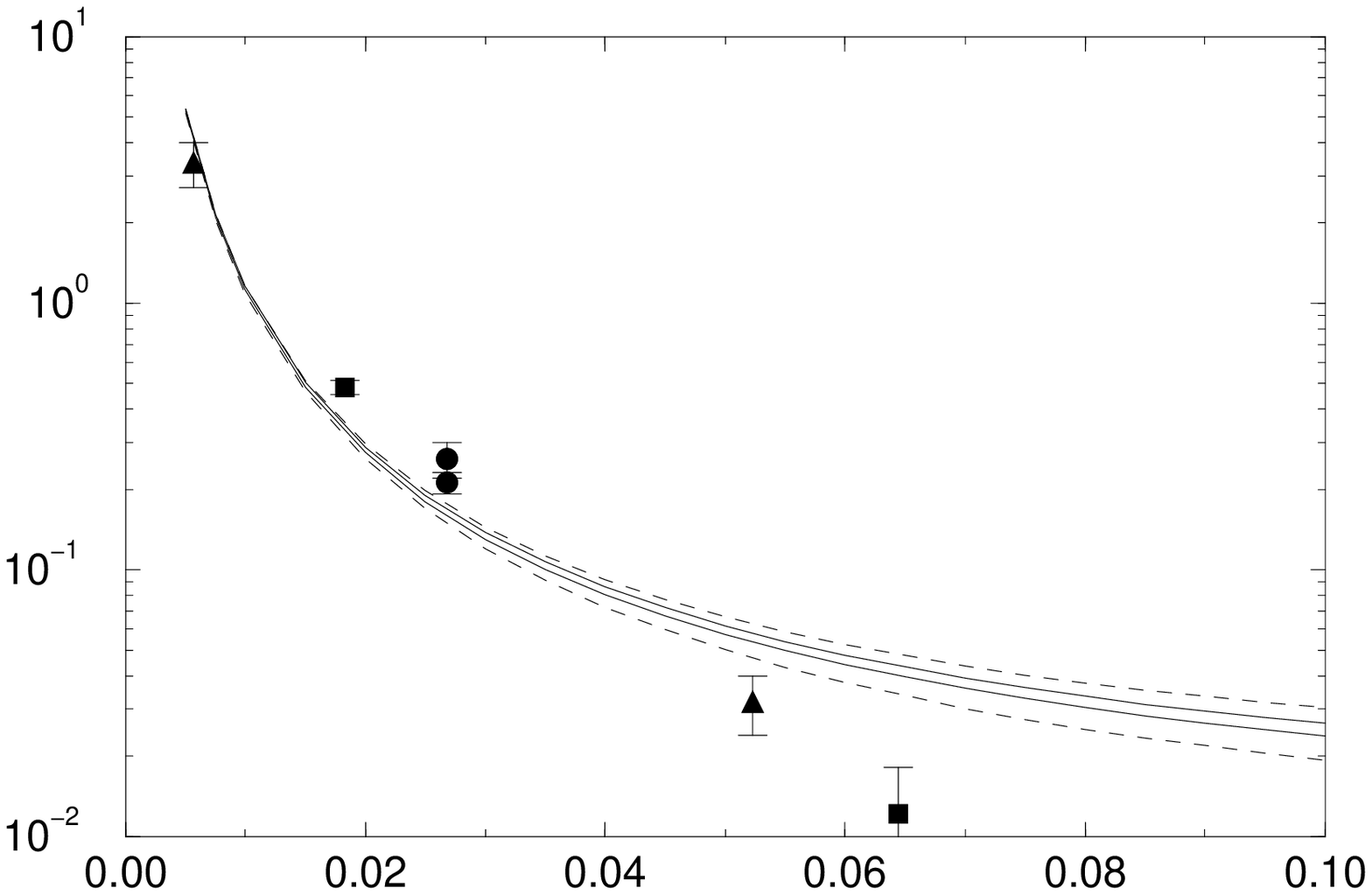}}
\put(270,0){$x\rightarrow$}
\put(60,140){$\frac{g_W^2\sigma}{(g_3^2)^3}\uparrow$}
\end{picture} 
\caption{(from ref. [15]). The perturbatively calculated 
interface tension $\sigma$ (including $Z$-factor effect and
gauge variations) vs.~$x$ compared to lattice data from 
ref. [10] (squares), ref. [16] (triangles) and 
ref. [17] (circles).}
\label{fig2} 
\end{figure}

The most secure way to deal with (\ref{2.1}) is to use it in lattice
calculations \cite{9,10}
to determine the critical temperature $T_c$ and the
interface tension and latent heat -- if we have a first-order PT.
An alternative treatment would be in the framework of Wilsonian
renormalization \cite{11}. $L^{3-dim}_{eff}$ of eq. (\ref{2.1}) is an
$x,y$-dependent superrenormalizable 3-dimensional Lagrangian with just
one scale $g^2_3$ and without fermions and thus can be handled
very safely in the lattice approach. The results of such lattice
calculations \cite{9,10} are:

(i) there is a first-order PT for $x\lsim 0.11$;
there is a second-order
PT at the endpoint \cite{12a}
and above $x=0.11$ one has a crossover -- no
PT anymore!\\

(ii) $v(T_c)/T_c=\varphi_{min}(T_c)/
T_c\gsim 1\quad {\rm for}\quad x\lsim 0.04$\\

(iii) Comparing the 2-loop perturbative expressions obtained
from (\ref{2.1}) with lattice results, there are deviations for $x\gsim
0.05$ in particular for the interface tension (fig. \ref{fig2}).\\

To protect a previously generated baryon asymmetry in a universe
with \linebreak $B-L=0$ from erasure by sphaleron transitions
$\sim \exp(-Av(T)/T)$ in a thermodynamic equilibrium
period inside the Higgs phase  one needs $v(T_c)/T_c\gsim 1$.
With $x=(1/8)m_H^2/m_W^2+c_{Pos}m^4_t/m^4_W$
where the second term alone is $>0.04$ for the observed top mass
$m_t$, this can never be achieved in the SM, independent of the
Higgs mass. Together with its CP-violating effects being smaller than 
needed for an asymmetry production, this prevents the SM to explain the
baryon asymmetry of the universe.

\section{Nonperturbative effects in the \protect\\ three-dimensional
electroweak potential}
\setcounter{equation}{0}

Lattice results give  a clear picture for the phase diagram
in the case of Lagrangian (\ref{2.1}). However, for some
questions -- e.g. sphaleron action, shape and action of
the critical bubble -- an explicit effective (coarse-grained)
action still would be useful. It is also very important to have
some (semi)analytic picture which tells us where one can trust
perturbation theory and where not. This will be particularly true
in the case of more complicated effective actions where lattice
results may not be available. Thus we shortly discuss such a model
\cite{12}.

\begin{figure}[t] 
\begin{picture}(100,100)
\put(150,0){\epsfxsize4cm \epsffile{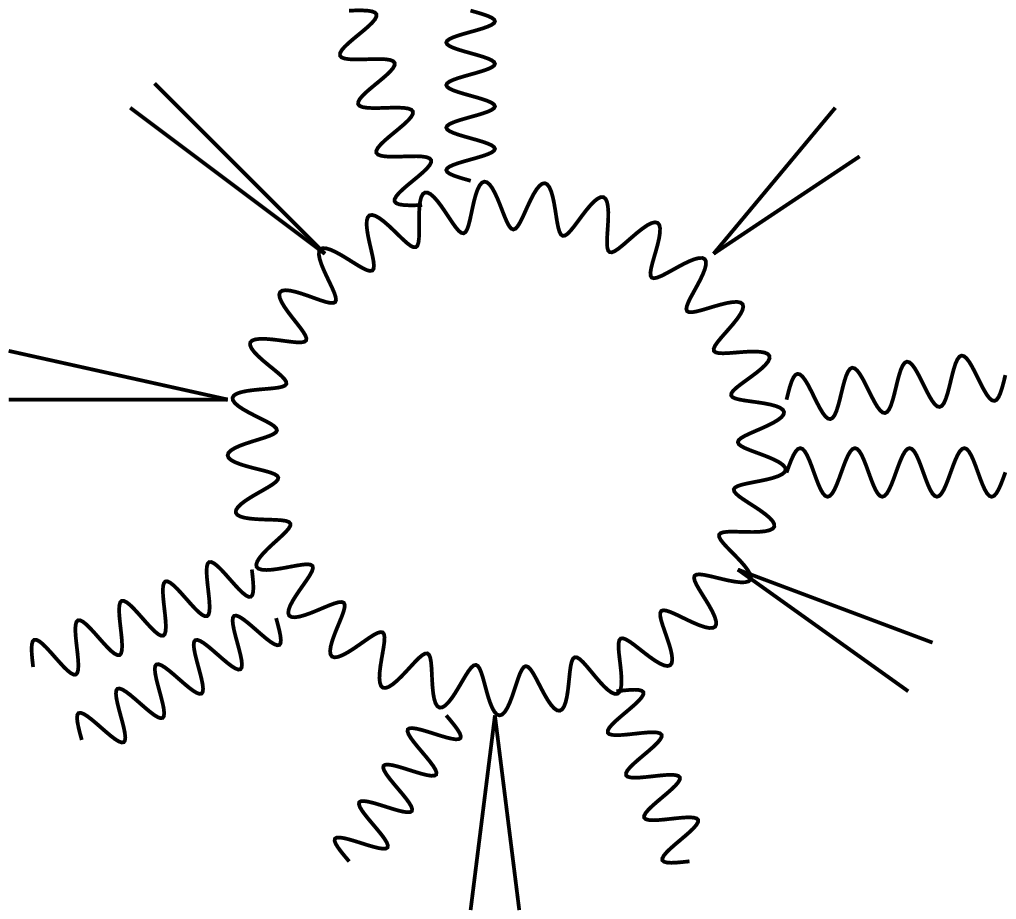}}
\put(150,26){x}
\put(156,17){x}
\put(185,98){x}
\put(199,100){x}
\put(185,5){x}
\put(223,5){x}
\put(257,58){x}
\put(257,47){x}
\end{picture} 
\caption{1-loop graph contributing to the potential $V(\varphi^2,<g_3^2F^2>)$.}
\label{fig3} 
\end{figure}

In the hot symmetric phase with background $\varphi=0$ the
Lagrangian (\ref{2.1}) describes a 3-dimensional QCD-type theory
with scalar Higgs ``quarks''. Lattice calculations \cite{10} show that
indeed in this phase static ``quarks'' experience a constant
string tension which furthermore is approximately equal
to that of pure SU(2)-Yang-Mills theory. This hints to a nonperturbative
dynamics dominated by ``W-gluons''. Also a spectrum of correlation masses
of gauge-invariant $H\bar H$ bound states and of $W$-glueballs
has been calculated on the lattice \cite{13}. The former is compatible
with a linearly rising potential in a relativistic bound state
model \cite{14}
(like that of Simonov in 4-dimensional QCD \cite{15}). There is only
a small mixing with the W-glueballs \cite{13}
in agreement with the suggestion
above that we have pure ``W-gluon''dynamics.

An interesting phenomenological description of the QCD vacuum is the
``stochastic vacuum model'' of Dosch and Simonov
\cite{16,17}. Its main virtue
is that it leads very naturally to the area law of confinement.
We have applied it to the 3-dimensional theory (\ref{2.1})
with an $SU(2)_W$ gauge group. Its main ingredient is a correlated
gauge field background with a purely Gaussian correlation
\be\label{3.1}
\ll g^2_3F^a_{i\kappa}(x')F^a_{i\kappa}(x)\gg =<g^2_3
F^2>D\left(\frac{(x-x')^2}{a^2}\right)\mbox{ }.\ee
This correlator is already simplified by  choice of a
coordinate gauge and by averaging over the tensor structure.
$<g^2_3F^2>$ is the normalization by the usual local gauge
field condensate and $D$ $(D(0)=1)$ is a form factor containing the
correlation length $a$. The correlator has been
tested in 3-dimensional lattice calculations
\cite{18} and the
correlation length was obtained as $a\sim1/0.73g^2_3
\sim2/m_{glueball}$. In ref. \cite{12} we presented strong indications 
that the $<g_3^2F^2>$ ground state is unstable (similar to the
Savvidy instability of QCD) for small Higgs vevs. Thus one 
obtains nonperturbative effects by a fluctuating gauge field 
background of type (\ref{3.1}).

\begin{figure}[t] 
\begin{picture}(200,160)
\put(-140,-20){\epsfxsize7cm \epsffile{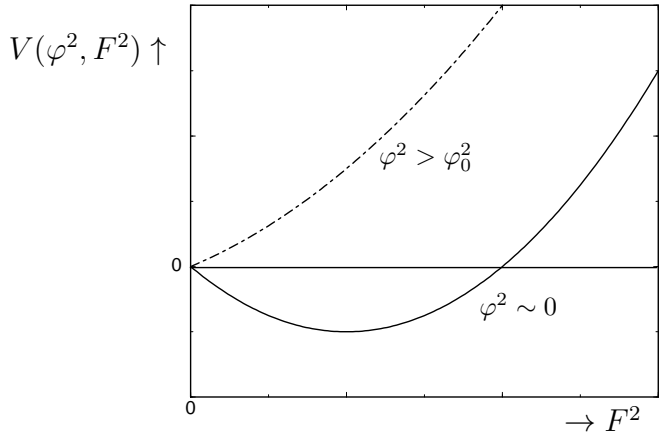}}
\put(260,-1){$\rightarrow F^2$}
\put(50,140){$V(\varphi^2,F^2)\uparrow$}
\put(228,43){\footnotesize{$\varphi^2\sim 0$}}
\put(190,100){\footnotesize{$\varphi^2>\varphi^2_0$}}
\end{picture} 
\caption{Sketch of the potential $V(\varphi^2,<g_3^2F^2>)$ in
$F^2$-direction for two different values of $\varphi^2$.}
\label{fig4} 
\end{figure}

\begin{figure}[p] 
\begin{picture}(200,180)
\put(-240,10){\epsfxsize7cm \epsffile{fig5a.eps}}
\put(-5,10){\epsfxsize7cm \epsffile{fig5b.eps}}
\put(325,20){ $m^2[g_3^4]$}
\put(175,170){a)}
\put(410,170){b)}
\put(90,20){ $p^2[g_3^4]$}
\put(32,170){$p^2$}
\put(277,120){$m^2_{\rm conf}$}
\put(266,66){$\tilde{S}_F$}
\put(33,55){$m^2_{\rm conf}$}
\put(98,80){$\tilde{S}_F$}
\end{picture} 
\caption{(from ref. [13]). $m_{\rm conf}^2(p^2,m^2)$ and $\tilde{S}_F(p^2,m^2)$ 
in units of $(g_3^2)^2$ plotted a) for $m^2=0$ and b) $p^2=0$.}
\label{SFM} 
\end{figure}

\begin{figure}[p] 
\begin{picture}(200,180)
\put(175,200){a)}
\put(-240,10){\epsfxsize7cm \epsffile{fig6.eps}}
\put(90,20){ $\varphi[g_3]$}
\put(121,75){$V_1^H$}
\put(50,145){$V_1^g$}
\put(113,160){$\varphi^3_g$}
\put(410,200){b)}
\put(-5,10){\epsfxsize7cm \epsffile{fig7.eps}}
\put(325,20){ $\varphi[g_3]$}
\put(230,114){$\frac{V}{(g_3^2)^3}\uparrow$}
\put(344,160){$x_3$}
\put(373,133){$x_2$}
\put(403,100){$x_1$}
\end{picture} 
\caption{(from ref. [13]). a) $V_1^g(\varphi)$, $V_1^H(\varphi)$ compared 
to the perturbative $\varphi^3_g$ term (in units of $(g_3^2)^3$). 
b) Fading away of the first-order phase transition
with increasing $x=\frac{\lambda}{g_3^2}$, where $x_1=0.06$,
$x_2=0.08$ and $x_3=0.11$.}
\label{fig5} 
\end{figure}

One can estimate the effect of such a background on the W-boson
(and ghost) loop leading to the 1-loop effective potential
$V(\varphi^2,<g^2_3F^2>)$ (fig. \ref{fig3}). We found \cite{12} two
contributions to a momentum-dependent effective (``magnetic'') mass:

(i) an IR regulator mass $m^2_{conf}(p^2,\varphi^2,<g^2_3
F^2>)$ of gauge bosons and ghosts due to
the string tension (area law) which
cures the IR problems of perturbation theory.

(ii) a negative effective (mass)$^2$ for the W-bosons $-\tilde S_F
(p^2,\varphi^2,<g^2_3F^2>)$ due to spin-spin forces which becomes
important for larger $p^2$ (``paramagnetism'') and does not
spoil the nice IR properties of $m^2_{conf}$.
If we introduce these masses in the 1-loop action (gauge boson loop)
it has roughly the form \pagebreak
\bear\label{3.2}
V(\varphi^2,<g^2_3F^2>)&\sim&...\int\frac{d^3p}{(2\pi)^3}log[p^2
+\frac{1}{4}g^2_3\varphi^2+m^2_{conf}
(p^2,\varphi^2,<g^2_3F^2>)\nonumber\\
&&-\tilde S_F(p^2,\varphi^2,<g^2_3F^2>)] \mbox{ }.
\ear
(This has to be corrected \cite{12} for combinatorics and also has to be
renormalized). Both masses depend on $<g^2_3F^2>$. Expanding (\ref{3.2})
in first-order in 
$<g^2_3F^2>$ the spin-spin force in
$-\tilde S_F$ produces the well-known negative
$F^2$-term destabilizing the $F^2=0$ vacuum. Adding the tree
$\frac{1}{4}F^2$ we can obtain an effective potential sketched in
fig. \ref{fig4} stabilized  at some value $F^2\not=0$ by confinement forces.
This is a 1-loop calculation and the masses
$m^2_{conf}$ and $-\tilde S_F$ are
determined only roughly (in lack of lattice data support). Thus we have
only a qualitative picture. To proceed, we fixed $<g^2_3F^2>$ at
the minimum by a relation to the lattice string tension.

Fig. \ref{SFM} shows the qualitative form of $m^2_{conf}(p^2,\varphi^2)$ and
of $\tilde{S}_F(p^2,\varphi^2)$ and fig. 6a the modified
``$\varphi^{3}$''-term
corresponding to (\ref{3.2}). Fig. 6b presents the new 1-loop
potential at the critical temperature at various $x$-values,
and one can see the first-order PT fading away. One can
also evaluate the interface tension (table) and determine roughly
the crossover point by postulating that the effective $\varphi^2$
and $\varphi^4$ vanish at this (conformal) point with a
second-order PT.

\vspace{0.5cm}
\centerline{
\begin{tabular}{|c||c|c|} \hline
$x$ & $\sigma$ & $\sigma_{perturbative}$ \\ \hline \hline
0.06 & 0.016 & 0.013 \\ \hline
0.08 & 0.004 & 0.007 \\ \hline
0.11 & 0 & 0.004 \\ \hline
\end{tabular} }
\vspace{0.5cm}

We should stress again that this picture of nonperturbative
effects is not really quantitative, in particular because
2-loop calculations  in a correlated gauge-field background are
(too) difficult. Still we might get an indication in which
direction nonperturbative contributions go.

\section{ The MSSM with a ``light'' stop}
\setcounter{equation}{0}

\begin{figure}[t] 
\begin{picture}(100,100)
\put(150,0){\epsfxsize4cm \epsffile{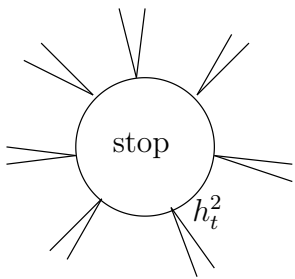}}
\put(193,50){stop}
\put(223,24){$h_t^2$}
\end{picture} 
\caption{1-loop stop contribution to the effective potential.}
\label{fig6} 
\end{figure}

Searching for modifications of the electroweak theory in
order to obtain a strongly first-order PT, one
faces  the by now sufficiently known situation
that the success of the standard model is both blessing \underbar{and}
burden. We do not have experimental hints which way to go.
Supersymmetric theories have the well-known theoretical
advantages. From a practical point of view all one needs
for a strongly first-order PT is the strengthening of the
``$\varphi^3$''-term in the effective potential due to bosonic
exchange in the loop. Thus one needs further bosons with a strong
coupling to the Higgs. SUSY models have a host of new bosons
in the superpartner sector. In particular the $s$-top particles
have a particularly strong Yukawa coupling $h_t$ if the Higgs vev $<v_2>$
of the Higgs coupling to the top
$(m_{top}=h_t<v_2>)$ is not very large, i.e. if
$\tan\beta=v_2/v_1$ is not large. The superpartner of the right-handed
top, the $stop_R$, does not have
$SU(2)_W$ interactions, and thus is particularly flexible in its
allowed mass (no $\rho$-parameter problem). As proposed in ref. \cite{19,20},
its exchange (fig. \ref{fig6}) 
can enhance the PT significantly if its
mass $m^2_3$ in the symmetric phase (including $T^2$-plasma mass)
is small:
\be\label{4.1}
m^2_3=m^2_0+cT^2\ee
where $m^2_0$ is the SUSY-breaking scalar mass of the $stop_R$.
The $T=0$ mass of the $stop_R$ is
\be\label{4.2}
m^2_{\tilde t}=m^2_0+m^2_t\ee
and is not much larger than the top mass for small positive
$m^2_0$. There might be a nonuniversal SUSY mass breaking
at the GUT scale necessary for very small $m_0^2$ though the
stop mass$^2$ is naturally lowered by renormalization flow.

If the $stop_R$ and one heavy combination of Higgses
is integrated out, one is led again to a Lagrangian
of the form (\ref{2.1}), but now with an $x=\lambda_3/g^2_3$ value
much smaller than in the SM (being bosonic the stop contributes opposite to the top!) 
allowing for $m_H\lsim 75$ GeV for a strongly
first-order PT with $v(T_c)/T_c\geq1$ \cite{22}-\cite{25}.

One can also ask \cite{19} for $stop_R$ masses smaller than the top mass
taking $m^2_0=-\tilde m^2_0$ negative in
(\ref{4.1}), (\ref{4.2}). The $stop_R$ than should not be fully
(also zero modes) integrated out, but kept in the effective 3-dimensional
zero mode action together with the light Higgs fields. If one
assumes that the CP-odd Higgs $A_0$ meson surviving spontaneous breaking
is rather heavy $(\gsim 300$ GeV), there is a heavy Higgs sector to
be integrated out, and just as above one Higgs field remains. We thus have
to consider a Lagrangian \cite{23}
\bear\label{4.3}
L^{3-dim}_{eff}&=&L^{3-dim}_{eff}(Higgs)\nonumber\\
&&+\frac{1}{4}G^A_{ij}G^A_{ij}+(D^s_iU)^+(D_i^sU)+
m_{U_3}^2U^+U\nonumber\\
&&+\lambda_{U_3}(U^+U)^2+\gamma_3(\phi^+_3\phi_3)(U^+U) \mbox{ }.\ear
The $T$-dependent parameters are obtained by integrating
out all non-zero modes and all heavy particles like in (\ref{2.1}),
which is the first part of the Lagrangian (\ref{4.3}). Thus one
has to specify the field content and the SUSY-breaking
parameters of the model. The simplest choice is the minimal
supersymmetric standard model (MSSM) \cite{21}
without universality for the top
scalar SUSY-breaking masses. The partner of the left-handed top
with a SUSY-breaking mass $m^2_Q$ should be heavy in order not to
contribute too much to $\Delta\rho$.

\begin{figure}[p] 
\begin{picture}(200,120)
\put(95,-10){\epsfxsize7cm \epsffile{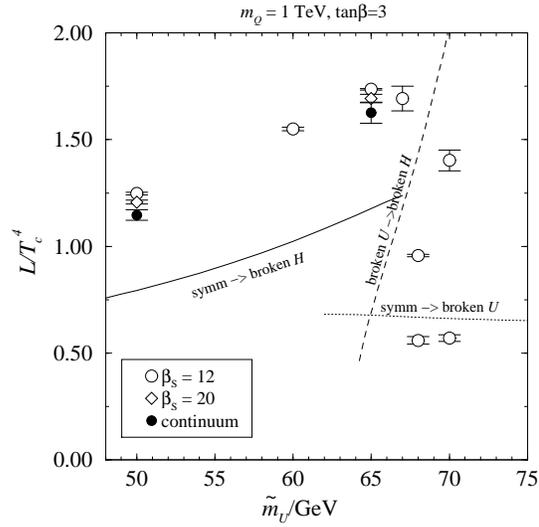}}
\end{picture} 
\caption{Latent heat at $T_c$ in dependence of mass parameter
$\tilde{m}_U$ calculated on the lattice in ref. [33] compared to
the analytic results of ref. [31].}
\label{fig7a} 
\end{figure}

\begin{figure}[p] 
\begin{picture}(200,120)
\put(-135,-10){\epsfxsize7cm \epsffile{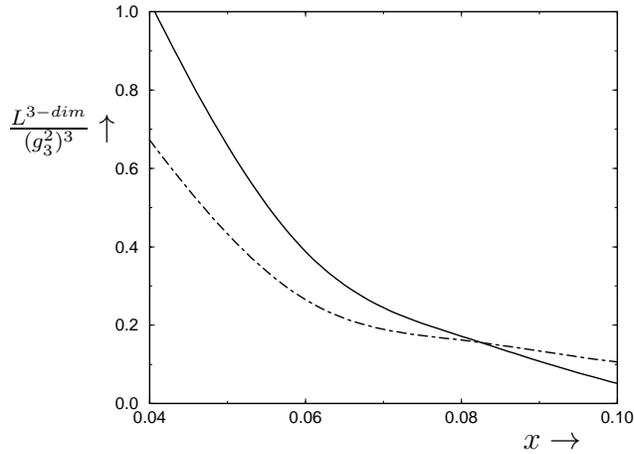}}
\put(265,5){$x\rightarrow$}
\put(70,123){$\frac{L^{3-dim}}{(g_3^2)^3}\uparrow$}
\end{picture} 
\caption{Three dimensional latent heat in dependence of $x$ 
calculated from the potential (3.2) (full line) compared to 
the result of ordinary 1-loop perturbation theory 
(dashed-dotted line).}
\label{fig7b} 
\end{figure}

Two-loop calculations with (\ref{4.3}) have shown that one can
indeed obtain \linebreak $v(T_c)/T_c\gsim1$ even for lightest Higgs masses
as big as 105 GeV \cite{26}. The para- \linebreak meter 
space is enlarged \cite{27}
if one allows for $stop_R$-$stop_L$ mixing with a parameter
$\tilde A_t=A_t+\mu ...$ . (Both parameters $\mu$ and
$A_t$ are important in the discussion of CP-violations
in the wall.) Ref. \cite{27} uses an improved 4-dimensional
one-loop effective potential at high temperatures and still
agrees well with the special case considered in \cite{26}.

For large enough negative $m^2_U=-\tilde m^2_U$ one even
obtains \cite{26,27} a two-stage phase transition with an intermediate
stop condensate $<U^+U>$. This is only acceptable if the
transition rate which is rapidly decreasing  with increasing
$\tilde m^2_U$ still allows to return from the stop phase
to the Higgs phase. In the former phase one has a situation
analogous to the Higgs phase, in particular
massive $SU(3)$ gauge bosons.

Recent lattice calculations confirm the perturbative results
surprisingly well \cite{28} (fig. \ref{fig7a}) --
though there are also significant
deviations. In particular the PT turned out to be more strongly
first order -- the latent heat and $v(T_c)/T_c$ are larger
than in the perturbative result. We can understand this effect
qualitatively with our model for nonperturbative contributions:
The effective $x$-value in the Higgs part of (\ref{4.3}) is much
smaller than in the standard model and for these values (fig. \ref{fig7b})
the latent heat and $v(T_c)/T_c$ are both increased compared to
pure perturbation theory. The important additional graphs coming
from Lagrangian (\ref{4.3}) mostly involve $SU(3)$ gluons
and the $stop_R$ both of which do not have $SU(2)_W$ interaction,
and hence also no nonperturbative effects on
this scale\footnote{This remark may be used to consider a hybrid
model combined out of lattice and perturbative calculations.}.

\section{NMSSM with a strongly
first-order phase transition}
\setcounter{equation}{0}

In the effective electroweak potential near the critical
temperature a term of type $-\varphi^3$ triggers a first-order
PT. Up to now we discussed the generation of such
terms in 1-loop order of perturbation theory. There is also
the possibility to obtain it already on the tree level. An
$SU(2)_W$-invariant third-order polynomial term in the
potential cannot just contain the Higgs(es). Thus one has
to enlarge the field content of the SM and also of the MSSM
in the case of a supersymmetric theory. The simplest extension
of the MSSM, the ``next to minimal model'' NMSSM \cite{29,29a}, contains
a further superfield $S$, which is a gauge singlet, 
in an additional piece of the superpotential
\be\label{5.1}
g^S=\lambda SH_1H_2-\frac{k}{3} S^3.\ee
The soft SUSY breaking term
\be\label{5.2}
V^S=A_\lambda \lambda SH_1H_2-\frac{k}{3}A_kS^3\ee
has the desired ``$\varphi^3$'' form \cite{30} if $S$ is on the same
level as the $H_i$. The superpotential (\ref{5.1})
has the virtue to avoid the $\mu$-term $g^\mu=\mu H_1H_2$
with its fine-tuning problem because
this term automatically arises after the singlet field aquires a vev. However,
because of its $Z_3$ symmetry it suffers from  the well-known
domain wall problem \cite{31}.
It turns out that the NMSSM with just
(\ref{5.1}) and (\ref{5.2}) besides having the domain wall problem
also is unable to produce a phase transition in $<S>$ and
$<H>$ simultaneously, which requires $<S>$ and $<H>$ to be
of the same order of magnitude. With a very large $<S>$\footnote{It
was shown in ref. \cite{29a} that in the case of 
universal soft SUSY breaking at the GUT scale the singlet vev has to
be larger than 1 TeV to prevent the appearence of not observed light particles 
in the spectrum. Similar results were obtained for certain non-universal
soft terms \cite{29b}.}
one would first obtain a PT in $<S>$ and afterwards the ordinary MSSM
PT in some Higgs field combination, which is not what we want.
We thus as in ref. \cite{32} choose the superpotential\footnote{We
are aware that removing the cosmologically problematic
$Z_3$ symmetry after inclusion of non-renormalizable interactions
may reintroduce quadratically divergent singlet
tadpoles which can destabilize the electroweak scale \cite{32a}. 
But there are some suggestions in the literature how to prevent 
these potentially dangerous diagrams \cite{32b}  
or even use them in a constructive way as a tool for model building \cite{32c,33}.} 
\be\label{5.3}
g=g^S+\mu H_1H_2-rS.\ee
Different from work \cite{34} more than a decade  ago we keep the full
parameter space of the model only restricted by universal
SUSY breaking at the GUT scale. In the latter we differ
from ref. \cite{32} where the parameters were fixed at the
electroweak scale without such a criterion. Besides the well-known
gauge couplings in the $D$-terms we then have the parameters
$\lambda, k,\mu,r$ in the superpotential and for the SUSY breaking
a universal scalar mass squared $m^2_0$, a common gaugino mass $M_0$, 
as well as an analytic mass  term $B_0$ for the Higgses and a 
universal trilinear scalar coup- \linebreak ling $A_0$ corresponding to the 
second and third power terms in the superpotential, respectively.

Besides the tree potential and 1-loop Coleman-Weinberg corrections
we include 1-loop plasma masses for the $H_i$ and $S$ fields and 
the 1-loop ``$\varphi^3$''
terms discussed in previous chapters which, however, now in
general are small compared to the tree term (\ref{5.2}). The most 
important finite temperature contributions come from the top quark
and the gauge bosons, but in some parts of the parameter space
the stops, charginos and neutralinos may become rather light
and therefore are also included in the effective potential $V_T(H_1,H_2,S)$.

Having at hand the potential we are  interested in, a rather natural
procedure would be as follows: (Randomly) choose a set of the GUT scale 
parameters listed above. Then use the (1-loop) renormalization
group equations \cite{35} to evolve the parameters down to the 
weak scale and minimize the T=0 effective potential in order to
study the electroweak symmetry breaking. Of course, to reproduce 
the physical Z-boson mass $M_Z$ a rescaling of all the (unknown) 
dimensionful parameters is necessary. But 
in the very most number of cases after this rescaling 
there appear some unobserved light 
particles in the spectrum, so one has to try the 
next set of parameters and this whole ``shot-gun'' procedure is 
very inefficient. 

Instead we fix the T=0 electroweak minimum determined by $M_Z$, 
$\tan\beta=v_2/v_1$ and  $<S>$ in addition to the parameters 
$\lambda,k,m_0^2,M_0$,$A_0$ while $\mu$,$r$,$B_0$ 
remain unspecified. The important thing is that the latter
do not enter the 1-loop renormali- \linebreak zation group equations for 
$\lambda$, $k$ and the
soft parameters with exception of $B$ so we can calculate all 
parameters of the effective potential at the weak scale except
$\mu$, $r$ and $B$ which we determine by applying the 
minimization conditions 
\[\frac{\partial V_{T=0}(H_1,H_2,S)}{\partial H_i}=0 \quad, \quad\quad\quad
\frac{\partial V_{T=0}(H_1,H_2,S)}{\partial S}=0 \quad.\]
Because of the complicated 1-loop 
corrections these equations cannot be solved analytically, but an
iterative numerical solution taking the tree level solution as
starting values is possible.  Of course, whether the postulated
minimum $(M_Z,\tan\beta,<S>)$ is indeed the global minimum has
to be checked explicitly and constrains the para- \linebreak 
meter space of the model.
Using this procedure we are left with the seven parameters\footnote
{Additionally, we require the top quark mass $M_{top}=175$ GeV 
which allows us to fix the top Yukawa coupling as a function of 
$\tan\beta$. All the other Yukawa couplings are neglected which 
is only justified in the regime $\tan\beta\lsim 10$.}
\[\tan\beta,<S>,\lambda,k,m_0^2,M_0,A_0\]
which still contain a lot of freedom. Fortunately, not all parameters
are equally important with respect to the strength of the PT: 
Of most interest are the gaugino mass $M_0$ and the trilinear
scalar coupling $A_0$, as they determine the coefficients 
$A_{\lambda}$ and $A_k$ of the  ``$\varphi^3$''-terms in eq.
(\ref{5.2}). Therefore we will study the plane of these parameters
while keeping the others fixed. To maximize the lightest CP-even
Higgs mass $M_h$ $\tan\beta$ should be taken large while
$\lambda$ should be kept small. As stated above, a strong
PT can only be expected, if $<S>\sim M_Z$ which requires
$k$ to be not too small because of $<S>\sim \frac{A_k}{k}$.
The remaining parameter $m_0^2$ only influences the masses of
the additional Higgs bosons which we have chosen heavy.
 
\begin{figure}[t] 
\begin{picture}(120,190)
\put(-185,-20){\epsfxsize8cm \epsffile{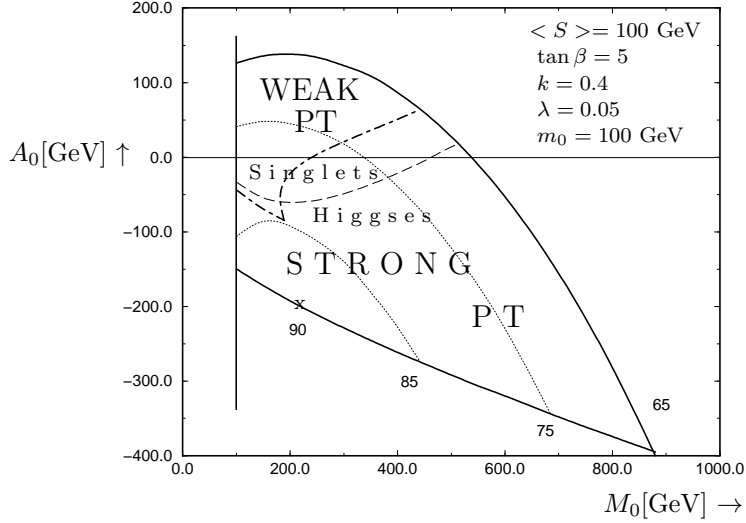}}
\put(270,-5){\footnotesize{$M_0$[GeV] $\rightarrow$}}
\put(45,128){\footnotesize{$A_0$[GeV] $\uparrow$}}
\put(153,72){\scriptsize{x}}
\put(242,175){\scriptsize{$<S>=100$ GeV}}
\put(245,165){\scriptsize{$\tan\beta=5$}}
\put(245,155){\scriptsize{$k=0.4$}}
\put(245,145){\scriptsize{$\lambda=0.05$}}
\put(245,135){\scriptsize{$m_0=100$} GeV}
\put(140,151){WEAK}
\put(153,139){PT}
\put(150,85){S T R O N G}
\put(220,65){P T}
\put(136,121){\scriptsize{S i n g l e t s}}
\put(160,105){\scriptsize{H i g g s e s}}
\end{picture} 
\caption{Scan of the $M_0$--$A_0$ plane where the remaining
parameters are fixed. The full line surrounds the phenomenologically
viable part of the parameter space. The dotted lines are curves of
constant lightest Higgs mass (75 and 85 GeV). The dashed line
indicates the region where the lightest Higgs is predominantely
a singlet. The dashed-dotted line separates the regions of strong 
($v_c/T_c\gsim 1$) and weak PT.}
\label{fig8} 
\end{figure}

An example of a scan in the $M_0$--$A_0$ plane is shown
in fig. (\ref{fig8}) where we fixed the remaining parameters
according to the remarks before as $<S>$=100 GeV,
$\tan\beta$=5, $\lambda$=0.05, $k=$0.4 and $m_0$=200 GeV.
There are several constraints on the parameter space: First of all,
the minimum postulated in the elimination procedure discussed
above has to be the global minimum which leads to the lower
bound on $A_0$ in fig. (\ref{fig8}). To prevent the appearence of
a chargino with mass smaller than 80 GeV the gaugino mass
$M_0$ has to be larger than 100 GeV corresponding to the
vertical line in the plot. Finally, we require the lightest Higgs
mass $M_h$ to be larger than 65 GeV which leads to the 
upper bound on $A_0$ in fig. (\ref{fig8})\footnote{Note that this
also implies an upper bound on the gaugino mass depending 
on the remaining parameters.}. Compared with the 
current LEP data on SM-like Higgs bosons this may seem 
to be a rather low value but one has to keep in mind that 
the lightest ``Higgs'' state in this model always has some
singlet component which even dominates in the region above 
the dashed line. Therefore the experimental constraints
on $M_h$ are somewhat relaxed.

In order to investigate the strength of the PT we determine
the critical temperature $T_c$ 
where there exist two degenerate minima in $V_T(H_1,H_2,S)$, 
a broken minimum with $<H_i>\neq 0$ and a symmetric one 
with $<H_i>=0$\footnote{The singlet vev is different from zero
even in the symmetric minimum.}. For the previously discussed
set of parameters the results are again summarized in fig. (\ref{fig8}).
There the dashed-dotted line separates the region with a
weak PT from the region where the baryon number washout
criterium $v_c/T_c\gsim 1$ is fulfilled. One clearly sees that 
{\em most} of the parameter space is indeed compatible with
electroweak baryogenesis. Interestingly enough, the region
where the Higgs mass is maximized ($M_h\sim 90$ GeV) 
is not excluded. Let us again stress that the situation 
drastically changes if we increase the singlet vev
to e.~g.~$<S>=300$ GeV while decreasing $k$ in order to
obtain similar values of $M_h$. Then only a small range
of values of $A_0$ just above its lower bound allows
a strong PT and most of the parameter space leads
to erasure of the baryon asymmetry. 

In the previous example the maximal value of the Higgs mass is 90 GeV
but one can reach much higher values. By choosing $\tan\beta$=10
$M_h$=100 GeV can be obtained and still $v_c/T_c\gsim 1$ can be
fulfilled. Increasing the singlet vev to e.~g.~$<S>=250$ GeV allows
the even larger value of $M_h$=115 GeV without violating the washout
criterium. But with larger $<S>$ the amount of fine-tuning 
of $A_0$ increases and there is the danger of metastability since
the PT requires thermal tunneling over a rather high tree barrier.

\section{Concluding remarks; baryogenesis in the SUSY
electroweak phase transition}

Having found a model and a set of parameters where the
electroweak PT is strong enough to avoid sphaleron
erasure in the Higgs phase of a previously generated baryon
asymmetry, we are just at the beginning and not at the end
of the story: One now has to develop a consistent picture
of baryogenesis \cite{45}-\cite{51}.

First one has to derive the shape of the bubble wall of the
critical bubble \cite{43,44}. The corresponding action
determines the transition probability and -- together with
the Hubble parameter -- the degree of supercooling and the
nucleation temperature. Similarly the friction generated by
scattering processes at the bubble wall determines the shape of a
stationary expanding bubble \cite{47}.

Furthermore, CP-violation in the bubble wall formed by
a spatially varying Higgs-field combination has to be
discussed \cite{48}-\cite{50}. Different from the SM in SUSY models 
the phases
of the Higgs-field couplings $\mu$ and $A_t$ cannot be defined
away. CP could be broken explicitly or spontaneously by
Higgs condensates with a phase. The latter might only happen
in some temperature interval. This is very attractive since
it relaxes dangerous upper bounds by an experimentally allowed
neutron electric dipole moment \cite{50} (one also needs some explicit
CP-breaking in order to remove a sign ambiguity in the
spontaneous breaking).

In variants of the favourite ``charge transport mechanism'' \cite{51}
the scattering (transmission/reflection) of the particles in the
plasma (most important are Higgsinos, gauginos, stop) on the bubble
wall generates a current of chiral charge diffusing in the
symmetric phase in front of the expanding bubble and is transmitted
into a $B+L$ asymmetry by the ``hot'' (unsuppressed)
sphaleron processes during the time before the Higgs phase bubble
takes over. The bubble wall in new calculations \cite{52}-\cite{57} is favoured
to be thick\footnote{In case of the NMSSM there may exist some parts 
of the parameter space where the wall turns out to be thin.} 
(i.e. bigger than the mean free path of the scattering
particles) and is moving slowly ($\sim c/10 $, ``deflagration'').
Recently it was stressed in the literature \cite{52,53,57} that one should
not separate scattering at the bubble wall and diffusion
in the symmetric phase, but deal with both simultaneously
in the framework of quantum Boltzmann equations. The outcome
is still controversely discussed.

Concluding we can say that the strength of the electroweak phase
transition in specific models like the one discussed here can
be determined reliably  using a mix of perturbative theory and
lattice work, supplemented by a qualitative
analytic picture. If the PT is predicted by perturbative calculations
to be strongly first-order, this is not changed by nonperturbative
effects. In the MSSM with a ``light'' $stop_R$
and in a broad parameter range of the NMSSM with $\mu\not=0$
a strongly first-order PT with $v(T_c)/T_c\gsim1$ is
possible even at Higgs masses as big as 100 GeV (and even
higher?).

\section*{Acknowledgement}
We would like to thank P.~John, O.~Philipsen and M.~Reuter for 
useful discussions.
This work was supported in part by the
TMR network {\it Finite Temperature Phase Transitions in Particle 
Physics}, EU contract no. ERBFMRXCT97-0122.

\end{document}